\documentclass[12pt,aps,prd,a4paper,preprintnumbers,floatfix,nofootinbib,superscriptaddress,notitlepage]{revtex4}
\pdfoutput=1
\usepackage[english]{babel}
\usepackage{amsmath}
\usepackage{graphicx}
\usepackage{color}
\usepackage{mathrsfs}   
\usepackage{amssymb}
\usepackage[normalem]{ulem} 
\usepackage{dcolumn}
\usepackage{bm}
\usepackage{graphicx}
\usepackage[sort&compress]{natbib}
\usepackage{multirow}
\usepackage{rotating}
\usepackage{epstopdf}
\usepackage{diagbox}
\usepackage{booktabs}
\usepackage[utf8]{inputenc}
\usepackage{float}
\usepackage{xcolor}
\usepackage{mathtools}
\usepackage{bbm}
\usepackage{xspace}
\usepackage{booktabs}

\usepackage[colorlinks=true,linkcolor=redLinks,citecolor=greenLinks,urlcolor=redLinks, pdfborder={0 0 0},hyperfootnotes=false]{hyperref}

\newcommand {\be} {\begin{equation}}
\newcommand {\ee} {\end{equation}}

\definecolor{greenLinks}{rgb}{0, 0.6, 0} 
\definecolor{blueLinks}{rgb}{0, 0, 0.6}
\definecolor{redLinks}{rgb}{0.6, 0, 0}
\definecolor{tempText}{rgb}{0.55, 0.10,0.67}
\definecolor{eprintLinks}{rgb}{0.4, 0.4, 0.4}
\definecolor{journalLinks}{rgb}{0.6, 0, 0}

\def\21{$\mathrm{SU(2)_L \otimes U(1)_Y}$ }
\def\31{$\mathrm{SU(3)_c \otimes U(1)_Q}$ }

\def\3211{$\mathrm{SU(3) \otimes SU(2)_L \otimes U(1)_R \otimes U(1)_{B-L}}$ }
\def\321{$\mathrm{SU(3) \otimes SU(2) \otimes U(1)}$ }
\def\422{$\mathrm{SU(4) \otimes SU(2) \otimes SU(2)_R}$ }
\newcommand {\ignore}[1]{}

\usepackage{breqn}
\definecolor{darkgreen}{HTML}{109930}

\newcommand{\rep}[1]{\ensuremath\boldsymbol{#1}}
\newcommand{\crep}[1]{\ensuremath\bar{\boldsymbol{#1}}}
\newcommand{\Z}[1]{\ensuremath{\mathbbm{Z}_{#1}}} 
\newcommand{\D}[1]{\ensuremath{\mathrm{D}_{#1}}} 
\newcommand{\PS}[1]{\ensuremath{\mathrm{S}_{#1}}}
\newcommand{\SO}[1]{\ensuremath{\mathrm{SO}(#1)}}
\newcommand{\SU}[1]{\ensuremath{\mathrm{SU}(#1)}}

\newcommand{\e}[1]{\ensuremath{\mathrm{e}^{#1}}}
\newcommand{\I}{\mathrm{i}}

\newcommand{\T}{\ensuremath{\mathcal{T}_{13}}\xspace}

\newcommand{\AddrAHEP}{%
  AHEP Group, Institut de F\'isica Corpuscular --
  CSIC-Universitat de València, Parc Cient\'ific de Paterna.\\
 C/ Catedr\'atico Jos\'e Beltr\'an, 2 E-46980 Paterna (Valencia) - SPAIN}
 \newcommand{\AddrMPI}{Max-Planck-Institut f\"ur Kernphysik, Saupfercheckweg 1, 69117 Heidelberg, Germany}

\begin{document}
\title{\boldmath Asymmetric tri-bi-maximal mixing and residual symmetries}
\author{Salvador Centelles Chuli\'{a}}\email{salcen@ific.uv.es}
\affiliation{\AddrAHEP}
\author{Andreas Trautner}\email{trautner@mpi-hd.mpg.de}
\affiliation{\AddrMPI}

\begin{abstract}
Asymmetric tri-bi-maximal mixing is a recently proposed, grand unified theory (GUT) based, flavor mixing scheme. 
In it, the charged lepton mixing is fixed by the GUT connection to down-type quarks and a $\T$ flavor symmetry, 
while neutrino mixing is assumed to be tri-bi-maximal (TBM) with one additional free phase.
Here we show that this additional free phase can be fixed by the residual flavor and CP symmetries of the effective 
neutrino mass matrix. We discuss how those residual symmetries can be unified with $\T$ and identify the 
smallest possible unified flavor symmetries, namely $(\Z{13}\times\Z{13})\rtimes \D{12}$ and $(\Z{13}\times\Z{13})\rtimes \PS4$.
Sharp predictions are obtained for lepton mixing angles, CP violating phases and neutrinoless double beta decay.
\end{abstract}


\maketitle


\section{Asymmetric tri-bi-Maximal mixing}
\label{sec:intro}
Still there is no compelling theory of the observed masses and mixings of elementary fermions.
The only way to do away with the large number of apparently independent but correlated parameters is symmetry.
Especially promising is the concept of grand unification, where the 
disconnected quark and lepton Yukawa couplings of the Standard Model are consolidated. 
This immediately raises the question why the observed mixing in the quark and lepton sectors would be so different.
A straightforward explanation could be that a GUT induced similarity in the down-type quark and charged lepton mixings, 
in the spirit of ``quark-lepton complementarity'' \cite{Smirnov:2004ju,Minakata:2004xt,Raidal:2004iw}, is distorted by the subtle mass generation of neutrinos.
In this way, highly symmetric patterns for the neutrino mixing, such as TBM mixing \cite{Harrison:2002er,Harrison:2002kp}, would suffer a ``Cabbibo haze'' 
\cite{Datta:2005ci,Everett:2005ku,Kile:2013gla}, i.e.\ corrections stemming from the charged fermion sector \cite{Giunti:2002ye,Frampton:2004ud,Goswami:2009yy}.
This could reconcile a high symmetry in neutrino mixing with the somewhat off-symmetric observed values, especially for the large $\theta_{13}$.

\begin{figure}[t]
 \centering
  \includegraphics[scale=0.72]{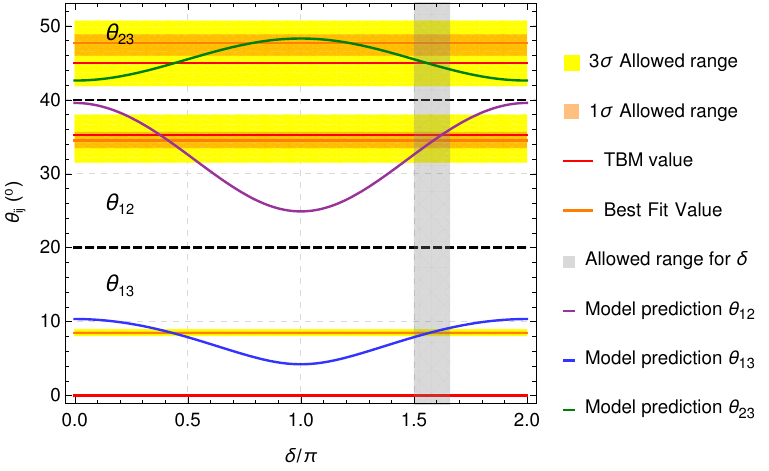}
  \includegraphics[scale=0.53]{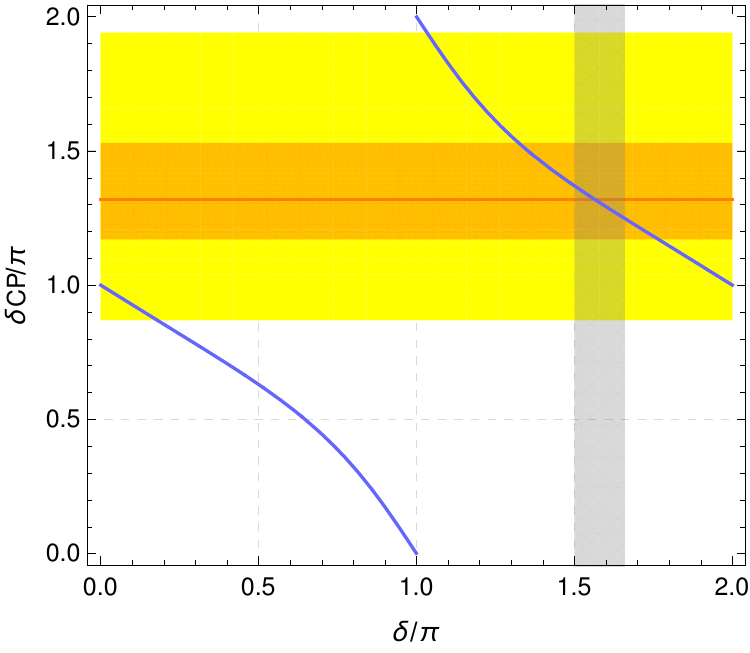}
      \caption{\label{fig:correlations}
      Predictions of asymmetric tri-bi-maximal mixing for lepton mixing angles (left) and Dirac CP violating phase (right) as a function of the internal parameter $\delta$.
      Experimentally allowed $1(3)\sigma$ regions are shown in orange(yellow). The allowed range for $\delta$ is shaded in gray.}
\end{figure}

However, pure CKM-type mixing for the charged leptons together with TBM
for the neutrino sector is observationally excluded. In fact, even for less symmetric patterns 
it was found that symmetric textures for the charged-lepton Yukawa matrix do not work~\cite{Kile:2014kya}.
It was therefore suggested to adopt a certain \textit{asymmetric} texture for the charged lepton sector~\cite{Rahat:2018sgs}.
In a concrete realization, the charged left-handed lepton mixing matrix then was found to be of the form~\cite{Rahat:2018sgs,Perez:2019aqq}
\begin{equation} \label{eq:Ucl}
U_{c\ell} ~=~ \left( \begin{matrix}
                1-\left( \frac{1}{18}+ \frac{2}{9A^2} \right) \lambda^2 & \frac{1}{3} \lambda & \frac{2}{3A} \lambda \\
                -\frac{1}{3} \lambda & 1-\frac{1}{18} \lambda^2 & A \lambda^2 \\
                -\frac{2}{3A}\lambda & -\left(A+\frac{2}{9A}\right)\lambda^2 & 1-\frac{2}{9A^2} \lambda^2
               \end{matrix} \right) + \mathcal{O}(\lambda^3)\;,
\end{equation}
where $\lambda=0.22453(44)$ and $A=0.836(15)$ are the usual quark sector Wolfenstein parameters~\cite{Tanabashi:2018oca}.
Together with a ``complex-TBM'' (cTBM) neutrino mixing matrix, defined as
\begin{equation}\label{eq:deltaTBM}
 U_{\nu}~=~U_{\mathrm{TBM}\delta}~:=~ 
 \begin{pmatrix}
\sqrt{\frac{2}{3}}            & \frac{1}{\sqrt{3}}  & 0 \\
-\frac{1}{\sqrt{6}}                      & \frac{1}{\sqrt{3}}  & \frac{1}{\sqrt{2}} \\
\frac{\e{\I \delta}}{\sqrt{6}} & -\frac{\e{\I \delta}}{\sqrt{3}} & \frac{\e{\I \delta}}{\sqrt{2}}
\end{pmatrix}\;,
\end{equation}
the Pontecorvo-Maki-Nakagawa-Sakata~(PMNS) lepton mixing matrix would be given by
\begin{eqnarray}
 U_{\mathrm{lep}}~=~U_{c\ell}^{\dagger} \, U_\nu\;.
\end{eqnarray}
Since $U_{c\ell}$ is fixed by the GUT connection to quarks and the reproduction of the CKM matrix, no residual symmetry is allowed for the charged lepton mixing.
In contrast, neutral lepton mixing in this framework is less constrained and it is possible that residual symmetries are responsible 
for a highly symmetric form of $U_\nu$.
We will parametrize the PMNS matrix in terms of the standard mixing angles and CP violating phases~\cite{Tanabashi:2018oca}.
The only free parameter here is $\delta$, and analytical expressions for all angles and phases can be obtained 
(we list them in Appendix \ref{app:AE}). Experimentally the most well-measured is 
\begin{equation}\label{eq:th13}
 \sin{\theta_{13}} \, = \, \frac{\lambda}{3\sqrt{2}}\left| 1+ \frac{2 \e{\I \delta }}{A} \right| + O(\lambda^3)\;. \\
\end{equation}
Varying $\theta_{13}$ and the Wolfenstein parameters in their 
experimentally allowed $3(1)\sigma$ range \cite{deSalas:2017kay, globalfit, deSalas:2018bym} 
constrains $\delta$ to lie in the interval $\delta/\pi\in\left\{1.50(1.53), 1.66(1.60)\right\}$, see Fig.~\ref{fig:correlations}.
The degeneracy between the sign of $\delta$ and the mixing angles is broken by the recent best-fit value of 
$\delta_{CP}$ \cite{deSalas:2018bym,Esteban:2018azc}.

A complete model based on \SO{10} unification with a flavor symmetry\footnote{%
See ref.\ \cite{Ding:2011qt,Hartmann:2011pq,Kajiyama:2010sb,Hartmann:2011dn} for more details on this symmetry.} 
$\T\cong\Z{13}\rtimes\Z{3}$ was put forward in \cite{Perez:2019aqq} which successfully realizes the asymmetric texture \eqref{eq:Ucl}. There, cTBM mixing was taken as an assumption, while it was suggested that it may arise from residual symmetries. --- If this scheme of mixing indeed originated from a residual symmetry, one would expect $\delta$ to take a value that is ``quantized'' in rational fractions of $\pi$. The allowed values of such a quantization intimately depend on the respective residual symmetry group. Exactly because of that, as we will clarify below, some rational values of $\delta/\pi$ are more special than others. We list some of them together with their phenomenological predictions in Tab.~\ref{tab:benchmark}.

In the present note, we derive all possible residual symmetries that realize the cTBM matrix, 
and clarify how they can be merged into a unified flavor symmetry together with the \T-symmetry of the charged fermion sector. We finish by discussing the phenomenological predictions of the most symmetric scenarios.
\begin{table}
\begin{center}
\begin{tabular}{ccccc}
  \toprule[1pt]
  \hspace{0.2cm}  \hspace{0.2cm} & \hspace{0.2cm} $\sin^2 \theta_{12}$ \hspace{0.2cm} & \hspace{0.2cm} $\sin^2 \theta_{13}$ \hspace{0.2cm} &\hspace{0.2cm} $\sin^2 \theta_{23}$ \hspace{0.2cm} &\hspace{0.2cm} $ \delta_{CP}$ \hspace{0.2cm} \\
  \hline
  \hspace{0.2cm} Best fit (NO) \hspace{0.2cm} & \hspace{0.2cm}  $0.320$ \hspace{0.2cm} & \hspace{0.2cm}  $0.02160$ \hspace{0.2cm} &\hspace{0.2cm}  $0.547$ \hspace{0.2cm} & \hspace{0.2cm}  $1.32\pi$ \hspace{0.2cm} \\
  \hspace{0.2cm} $1\sigma$ range (NO) \hspace{0.2cm} & \hspace{0.2cm}  $0.304\div0.340$ \hspace{0.2cm} & \hspace{0.2cm}  $0.02091\div0.02243$ \hspace{0.2cm} &\hspace{0.2cm}  $0.517\div0.567$ \hspace{0.2cm} & \hspace{0.2cm}  $1.17\pi\div1.53\pi$ \hspace{0.2cm} \\
  \hspace{0.2cm} $3\sigma$ range (NO) \hspace{0.2cm} & \hspace{0.2cm}  $0.273\div0.379$ \hspace{0.2cm} & \hspace{0.2cm}  $0.0196\div0.0241$ \hspace{0.2cm} &\hspace{0.2cm}  $0.445\div0.599$ \hspace{0.2cm} & \hspace{0.2cm}  $0.87\pi\div1.94\pi$ \hspace{0.2cm} \\
 \hspace{0.2cm} $\delta = \mp9\pi/26$ \hspace{0.2cm} & \hspace{0.2cm} $0.343$ \hspace{0.2cm} & \hspace{0.2cm} $\bf{0.0251}$ \hspace{0.2cm} &\hspace{0.2cm} $\bf{0.485}$ \hspace{0.2cm} &\hspace{0.2cm} $\pm 1.25 \pi$ \hspace{0.2cm} \\
  \hspace{0.2cm} $\delta = \mp5\pi/13$ \hspace{0.2cm} & \hspace{0.2cm} $0.331$ \hspace{0.2cm} & \hspace{0.2cm} $\bf{0.0236}$ \hspace{0.2cm} &\hspace{0.2cm} $\bf{0.491}$ \hspace{0.2cm} &\hspace{0.2cm} $\pm 1.28 \pi$ \hspace{0.2cm} \\
 \hspace{0.2cm} $\delta = \mp11\pi/26$ \hspace{0.2cm} & \hspace{0.2cm} $0.317$ \hspace{0.2cm} & \hspace{0.2cm} $\bf{0.0220}$ \hspace{0.2cm} &\hspace{0.2cm} $\bf{0.496}$ \hspace{0.2cm} &\hspace{0.2cm} $\pm 1.31 \pi$ \hspace{0.2cm} \\
\hspace{0.2cm} $\delta = \mp6\pi/13$ \hspace{0.2cm} & \hspace{0.2cm} $0.304$ \hspace{0.2cm} & \hspace{0.2cm} $\bf{0.0204}$ \hspace{0.2cm} &\hspace{0.2cm} $\bf{0.502}$ \hspace{0.2cm} &\hspace{0.2cm} $\pm 1.34 \pi$ \hspace{0.2cm} \\
  \hspace{0.2cm} $\delta = \mp \pi/2$ \hspace{0.2cm} & \hspace{0.2cm} $0.289$ \hspace{0.2cm} & \hspace{0.2cm} $\bf{0.0188}$ \hspace{0.2cm} &\hspace{0.2cm} $\bf{0.508}$ \hspace{0.2cm} &\hspace{0.2cm} $\pm 1.36 \pi$ \hspace{0.2cm} \\
      \bottomrule[1pt]
  \end{tabular}
  \end{center}
  \caption{\label{tab:benchmark}
 Values of the lepton mixing parameters for some special values of the internal parameter $\delta$ for $A$ and $\lambda$ set to their best fit values, 
 together with the experimentally allowed ranges for normal mass ordering (NO). 
 The ranges and best fit points are very similar for inverted ordering (IO). 
 Note that $\delta = \pm 9\pi / 26$ and $\pm \pi/2$ are only allowed if the Wolfenstein parameters $A$ and $\lambda$ are also varied.}
\end{table}

\section{Symmetry origin of asymmetric TBM mixing}
\subsection{Residual symmetries of complex TBM mixing}
We start by discussing general residual symmetries of the effective lepton mass terms in the gauge basis 
\begin{equation}\label{eq:L}
 \mathcal{L}_{\mathrm{mass}}~=~-\bar{\ell}_{\mathrm{L}}\,M_\ell\,\ell_{\mathrm{R}} - \frac12\nu^{\mathrm{T}}_{\mathrm{L}}\,\mathcal{C}\,M_\nu\,\nu_{\mathrm{L}} + \mathrm{h.c.}\;,
\end{equation}
where $\ell$ denote charged leptons and $\nu$ the neutrinos.
The effective neutrino Majorana mass matrix $M_\nu$ is symmetric and $\mathcal{C}$ is the charge conjugation matrix.   
The unitary matrices $U_{\nu,c\ell}$ diagonalize the mass matrices 
\begin{align}\label{eq:diagonalization}
 U_\nu^{\mathrm{T}}\,M_\nu\,U_\nu~=~&\mathrm{diag}\left(m_1,m_2,m_3\right)~\;=:~D_\nu \;,\\
 U_{c\ell}^\dagger\left(M_\ell\,M_\ell^\dagger\right)U_{c\ell}~=~&\mathrm{diag}\left(m_e^2,m_\mu^2,m_\tau^2\right)~=:~D_\ell\;.
\end{align}
The diagonal mass matrices $D_i$ are all-real. 
The lepton mixing matrix is defined by 
\begin{equation}\label{eq:PMNS}
 U_{\mathrm{lep}}~:=~U_{c\ell}^\dagger\,U_\nu\;.
\end{equation}
A neutrino mixing matrix for $U_\nu$ of the form \eqref{eq:deltaTBM} can be enforced by residual symmetries.\footnote{%
These ``residual symmetries'' are not actual symmetries of the full electroweak Lagrangian
since they are, in particular, broken by the charged lepton mass terms.
Corrections due to this breaking are small scale-dependent effects that we will neglect for the purpose of our discussion.}
These could be global flavor symmetry transformations \cite{Ma:2005pd, Lam:2007qc, Blum:2007jz, Grimus:2009pg, Hernandez:2012ra}
\begin{equation}
 \nu_{(x,t)} \mapsto G\,\nu_{(x,t)}\;,
\end{equation}
with some $3\times3$ unitary matrices $G$, as well as general CP (GCP) transformations \cite{Feruglio:2012cw}\footnote{%
We have adjusted the unphysical global phase of the CP transformation to $\I$, such that 
$M_\nu$ is required to be real in a basis where CP is conserved with a matrix $X=\mathbbm{1}$.
}
\begin{equation}
 \nu_{(x,t)} \xmapsto{CP} \I\,X\,\gamma^0\,\mathcal{C}\,\bar{\nu}^{\mathrm{T}}_{(-x,t)}\;,
\end{equation}
with some $3\times3$ unitary matrices $X$. 
Invariance under these transformations requires 
\begin{equation}\label{eq:Ginvariance}
 G^{\mathrm{T}}\,M_\nu\,G~=~M_\nu\,\qquad\text{or}\qquad X^{\mathrm{T}}\,M_\nu\,X~=~{M^*_\nu}\;,
\end{equation}
respectively. 
We first focus on pure flavor symmetries and discuss details of possible additional residual GCP symmetries later.

Assume that $M_\nu$ in the gauge basis is invariant under a symmetry transformation with matrix $G$.
It follows that in the mass-diagonal basis \eqref{eq:diagonalization}, $D_\nu$ must be invariant under the basis transformed symmetry 
\begin{equation}
 E~:=~U_\nu^\dagger\,G\,U_\nu\;.
\end{equation}
The invariance condition \eqref{eq:Ginvariance} in this basis translates to  
\begin{equation}
 E\,D_\nu~=~D_\nu\,E^*\;.
\end{equation}
Assuming non-degenerate and non-zero neutrino masses, the unitary $E$ has to be diagonal and real which means it can only be one of the eight possibilities
(uncorrelated signs)
\begin{equation}\label{eq:E}
 E~=~\mathrm{diag}\left(\pm1,\pm1,\pm1\right)\;.
\end{equation}
Under the assumption of a specific form of the mixing matrix $U_\nu$, we can translate this back to the gauge basis as $G=U_\nu E U_\nu^\dagger$.
Assuming $U_\nu=U_{\mathrm{TBM}\delta}$ we find the eight possible residual symmetry transformations in the gauge basis:
\begin{align}\notag
 G^{\pm}_{1}~:=~&\pm\frac13
 \begin{pmatrix}
  -1              & 2            & -2\,\e{-\I\delta } \\
   2              & 2            & \e{-\I\delta } \\
  -2\,\e{\I\delta} & \e{\I\delta} & 2
 \end{pmatrix}\;,&
 G^{\pm}_{2}~:=~&\pm\frac13
 \begin{pmatrix}
  1               & -2           & 2\,\e{-\I\delta } \\
  -2              & 1            & 2\,\e{-\I\delta } \\
  2\,\e{\I\delta} & 2\,\e{\I\delta} & 1
 \end{pmatrix}\;,& \\\label{eq:G_gauge}
 G^{\pm}_{3}~:=~&\pm
 \begin{pmatrix}
  -1 & 0             & 0 \\
   0 & 0             & \e{-\I\delta} \\
   0 & \e{\I\delta } & 0 
 \end{pmatrix}\;,& 
  G^{\pm}_{4}~:=~&\pm
 \begin{pmatrix}
   1 & 0 & 0 \\
   0 & 1 & 0 \\
   0 & 0 & 1 
 \end{pmatrix}\,.& 
\end{align}
As already obvious from \eqref{eq:E}, and therefore completely independent of the phase $\delta$, these 
matrices make up for all $8$ elements of a group isomorphic to $\Z2\times\Z2\times\Z2$.
One of the $\Z2$ factors corresponds to the transformation with a global sign, $G=-\mathbbm{1}$, and is automatically a symmetry 
of the neutrino Majorana mass term in any basis, which is of course well known. 
Hence, this transformation does not impose any restrictions on $M_\nu$.
However, we still keep that factor explicit here as it could become meaningful once
we unify the residual symmetries of all left-handed leptons into a single group.
Furthermore, we note that the remaining $\Z2\times\Z2$ only requires two generators,
saying that many of the elements in \eqref{eq:G_gauge}
imply each other. For example, $G^{+}_1G^{+}_2=G^+_3$ \textit{etc.}

Consider now the action of possible residual GCP symmetries.
In close analogy to above we define 
\begin{equation}
P~:=~U_\nu^\dagger\,X\,U_\nu^*\;,
\end{equation}
upon which the invariance condition for a GCP transformation in the mass-diagonal basis reads
\begin{equation}
 P\,D_\nu~=~D_\nu\,P^*\;.
\end{equation}
Under the same assumptions as above this shows that also $P$ can only be one of the eight possibilities listed in \eqref{eq:E}.
Translating this back to the gauge basis via $X=U_\nu P U_\nu^{\mathrm{T}}$, while assuming the 
$U_{\mathrm{TBM}\delta}$ form for $U_\nu$, one finds the eight possible GCP symmetry transformations:
\begin{align}\notag
 X^{\pm}_{1}~:=~&\pm\frac13
 \begin{pmatrix}
  -1              & 2            & -2\,\e{\I\delta } \\
   2              & 2            & \e{\I\delta } \\
  -2\,\e{\I\delta} & \e{\I\delta} & 2\e{2\I\delta }
 \end{pmatrix}\;,&
 X^{\pm}_{2}~:=~&\pm\frac13
 \begin{pmatrix}
  1               & -2           & 2\,\e{\I\delta } \\
  -2              & 1            & 2\,\e{\I\delta } \\
  2\,\e{\I\delta} & 2\,\e{\I\delta} & \e{2\I\delta }
 \end{pmatrix}\;,& \\\label{eq:X_gauge}
 X^{\pm}_{3}~:=~&\pm
 \begin{pmatrix}
  -1 & 0             & 0 \\
   0 & 0             & \e{\I\delta} \\
   0 & \e{\I\delta } & 0 
 \end{pmatrix}\;,& 
  X^{\pm}_{4}~:=~&\pm
 \begin{pmatrix}
   1 & 0 & 0 \\
   0 & 1 & 0 \\
   0 & 0 & \e{2\I\delta } 
 \end{pmatrix}\,.& 
\end{align}
Together, these generate a group $\Z2^{\otimes4}$ irrespective of the choice of $\delta$. 
We stress that any combination of two GCP transformations, say with $X_i$ and $X_j$, 
induces a flavor symmetry transformation for the neutrinos with $G=-X_iX^*_j$.\footnote{%
The additional minus sign here is the usual minus sign that fermions pick up for (CP)$^2$ from their Lorentz structure.}
In this way, it is easy to confirm that the union of all GCP transformations, in fact, already contains (i.e.\ generates) all the eight 
previously stated flavor symmetries~\eqref{eq:G_gauge}.
The surplus \Z2 factor corresponds to a genuine CP transformation.

\subsection{Unification of complex TBM and \texorpdfstring{$\boldsymbol{\mathcal{T}_{13}}$}{T13}}
Above the electroweak scale we have $L=(\ell_{\mathrm{L}},\nu_{\mathrm{L}})$,
implying that the residual symmetries of left-handed charged leptons and left-handed neutrinos shall originate from a single group globally acting on $L$.
Therefore, we will now investigate how the symmetries that give rise to cTBM 
can be unified with the symmetry group \T, which gives rise to the asymmetric textures of the charged fermion Yukawa couplings
resulting in a mixing matrix \eqref{eq:Ucl}.

In the basis of \cite{Perez:2019aqq}, which is a gauge basis, 
the \T-symmetric transformation of the charged leptons is generated by the matrices
\begin{equation}
A~:=~
\begin{pmatrix}
 \rho & 0 & 0 \\
    0  & \rho^3 & 0 \\
   0   &    0    & \rho^9 
\end{pmatrix}\;,
 \qquad\text{and}\qquad
\tilde B~:=~
\begin{pmatrix}
 0 & 1 & 0\\
 0 & 0 & 1 \\
 1 &  0  & 0
\end{pmatrix}\;,
\end{equation}
where $\rho^{13}=1$. 
In the very same basis, possible residual symmetries of 
the neutrinos are generated by \eqref{eq:G_gauge} and \eqref{eq:X_gauge}.
A common unified flavor symmetry then should contain \textit{at least} the closure\footnote{%
Closure here means the matrix group generated by a set of matrix generators, and we will denote it by pointy brackets $\langle\dots\rangle$.}
of some of these elements.
Possible outcomes for the unified symmetry will depend on the number of elements that we ``add'' to \T,
and on the value of $\delta$. 

Before we start this analysis, let us make the following important observation:
We can do simultaneous basis transformations of all, charged and neutral, left-handed leptons.
These rotations are unphysical, in the sense that they cancel in the physical lepton mixing matrix.
Let us perform a simple simultaneous rephasing,
\begin{equation}
\left(\ell_{\mathrm{L}}',\nu_{\mathrm{L}}'\right)~=~\left(\tilde P\,\ell_{\mathrm{L}},\tilde P\,\nu_{\mathrm{L}}\right)\;,\qquad\text{with}\qquad \tilde P:=\mathrm{diag}(1,1,\e{-\I\delta})\;.
\end{equation}
This removes $\delta$ from the cTBM matrix, as well as from all of its residual symmetries 
(the outcome is the same as setting $\delta\rightarrow0$ in \eqref{eq:deltaTBM}, \eqref{eq:G_gauge}, and \eqref{eq:X_gauge}). 
This restores the usual TBM mixing, alongside its known residual symmetries! However,
$\delta$ now appears in $U_{c\ell}$, and $\tilde B$. The rephasing does not affect $A$.
Of course, the outcome for the unified group is independent of the chosen basis. 
In the re-phased basis, the only generator that contains $\delta$ is
\begin{equation}
B~:=~\tilde P\tilde B \tilde P^*~=~
\begin{pmatrix}
0  & 1 & 0 \\
 0 & 0 & \e{\I\delta} \\
 \e{-\I\delta} & 0  & 0
\end{pmatrix}\;,
\end{equation}
while $A$, and all the $G$'s and $X$'s take their values from above for $\delta\rightarrow0$.
The closure of $\left<A,B\right>$ is still \T, irrespective of $\delta$.
The overall unifying symmetry, however depends on the precise value of $\delta$ and which elements of $G_i$ and $X_i$ we pick in addition to $A$ and $B$.

\medskip

Let us briefly mention how family and GCP transformations are combined in practice.
We follow the procedure firstly outlined in \cite{Holthausen:2012dk} and collect spinors and their CP conjugates
in a single reducible representation, for example ${\nu_{\mathrm{L}}(x,t)}\oplus -\I\,\mathcal{C}\,{\nu_{\mathrm{L}}^*(-x,t)}$. We will suppress 
the space-time arguments in the following. Family and GCP transformations then act in a single matrix space,
\begin{equation}
\begin{pmatrix}
\nu_{\mathrm{L}} \\ -\I\,\mathcal{C}\,\nu_{\mathrm{L}}^* 
\end{pmatrix}
\xmapsto{G}
\begin{pmatrix}
G & 0 \\ 0 & G^*
\end{pmatrix}
\begin{pmatrix}
\nu_{\mathrm{L}} \\ -\I\,\mathcal{C}\,\nu_{\mathrm{L}}^* 
\end{pmatrix}\;,
\quad\text{or}\quad
\begin{pmatrix}
\nu_{\mathrm{L}} \\ -\I\,\mathcal{C}\,\nu_{\mathrm{L}}^* 
\end{pmatrix}
\xmapsto{GCP}
\begin{pmatrix}
0 & X \\ X^* & 0
\end{pmatrix}
\begin{pmatrix}
\nu_{\mathrm{L}} \\ -\I\,\mathcal{C}\,\nu_{\mathrm{L}}^* 
\end{pmatrix}\;,
\end{equation}
respectively. Family and GCP transformations then can be combined simply by matrix multiplication in this enlarged space.

\medskip

As a first result, note that sets $\left<A,G\right>$ involving $A$ and elements $G^{\pm}_{1,2}$ or $X^{\pm}_{1,2}$ \textit{never} close.
These, therefore, cannot be unified into a finite group irrespective of $\delta$.
They could, however be unified into continuous groups such as \SU3 or $\mathrm{G}_2$; this possibility is left for future work.

\medskip

On the other hand, elements of $G^{\pm}_{3,4}$ and $X^{\pm}_{3,4}$ form finite groups with $A$ and $B$.
We have listed these groups (or their orders, if too large) for certain phenomenologically allowed values of $\delta$ in table \ref{tab:groups}.
Coincidentally, groups involving $X^{\pm}_3$ often give rise to groups isomorphic to (or at least of the same size than) the extensions by $G^{\pm}_3$.
However, it is important to note that those are \textit{not the same} groups.
In particular, they will act differently on the fermions and their mass matrices:
For extensions with $G^{\pm}_{3}$, CP conservation is not required and fermions transform in a 
representation $\rep{3}\oplus\crep{3}$. If any GCP transformation is included, however, the fermions will transform
in an irreducible $\rep{6}$. Of course, the GCP transformations extend to neutral and charged lepton sectors likewise.
Thus, one may wonder whether or how those GCP transformations are consistent with 
\T which we observe to be a group of type I.\footnote{%
``Type I'' here refers to the classification of \cite{Chen:2014tpa}. 
Type I groups are groups that do not have a class-inverting automorphism. This prohibits transformations that simultaneously 
map all irreps of a group to their respective complex conjugate irreps. 
Type I groups have no basis with real Clebsch-Gordan coefficients.
The complex phases of the Clebsch-Gordan coefficients are known to be a symmetry-based source of explicit CP violation.}
Type I groups are inconsistent with CP transformations in generic settings.
However, this concrete model only features triplet irreps of \T (cf.~tab.~I of \cite{Perez:2019aqq})
for which consistent CP transformations are possible.
This situation would drastically change for a model that also contains non-trivial one-dimensional irreps of \T. 
In such a case, CP violation with calculable phases would be unavoidable. 
This would be an interesting starting point for further model building that we do not pursue here.

\medskip

It is straightforward to check that out of the $12$ independent real degrees of freedom in $M_\nu$, $G^{\pm}_{3}$ removes $4$, 
while $X^{\pm}_{3,4}$ individually each remove $6$. 
Taken together, $\langle G^{\pm}_{3}, X^{\pm}_{3},X^{\pm}_4\rangle$ remove $8$ degrees of freedom, leaving $4$. 
We will now focus on this most predictive case.
As is already clear from our discussions above, 
not all generators are needed to generate the maximal possible group $G_{\mathrm{max}}:=\langle A,B,G^{\pm}_{3}, X^{\pm}_{3},X^{\pm}_{4}\rangle$. 
Taking only $\langle G^{\pm}_{3},X^{\pm}_{3}\rangle$, $\langle G^{\pm}_{3},X^{\pm}_{4}\rangle$, 
or $\langle X^{\pm}_{3},X^{\pm}_{4}\rangle$ in addition to $A$ and $B$ is already sufficient to generate $G_{\mathrm{max}}$.
We have ignored the transformations $G^{\pm}_{4}$ here, because they do not lead to constraints on the neutrino mass matrix. 
Nevertheless, we remark that the according unphysical global \Z2 factor (generated by $G^{-}_{4}$) will unavoidably be part of the total flavor symmetry group if any CP transformations are included in the symmetry.
\begin{table}[t]
\begin{center}
\begin{tabular}{cccccc}
   \toprule[1pt]
   \hspace{0.2cm} $\pm\delta/\pi$ \hspace{0.2cm} & \hspace{0.2cm} $\langle A,B,G^{\pm}_{3}\rangle$ \hspace{0.2cm} & \hspace{0.2cm} $\langle A,B,X^{\pm}_{3}\rangle$ \hspace{0.2cm} & \hspace{0.2cm} $\langle A,B,G^{\pm}_{3} , X^{\pm}_{3},X^{\pm}_{4}\rangle$ \hspace{0.2cm} \\
   \cmidrule[0.5pt](l{.5em}r{.5em}){1-1} \cmidrule[0.5pt](l{.5em}r{.5em}){2-3} \cmidrule[0.5pt](l{.5em}r{.5em}){4-4}
   $0$  & $(\Z{13}\times\Z{13})\rtimes\PS4$  & \hspace{0.2cm} $(\Z{13}\times\Z{13})\rtimes\PS4$ \hspace{0.2cm} & $(\Z{13}\times\Z{13})\rtimes(\PS4\times\Z2)$\\
   $1/2$  &  $(\Z{13}\times\Z{13})\rtimes\PS3$ & $(\Z{13}\times\Z{13})\rtimes\PS3$ &  $(\Z{13}\times\Z{13})\rtimes \D{12}$\\
   \cmidrule[0.2pt](l{.5em}r{.5em}){1-4} 
   $2/5$ &  $\mathcal{O}(101{,}400)$ &  $\mathcal{O}(101{,}400)$ & $\mathcal{O}(202{,}800)$\\
   $3/7$ &  $\mathcal{O}(198{,}744)$ & $\mathcal{O}(198{,}744)$ & $\mathcal{O}(397{,}488)$ \\
   $3/8$ &  $\mathcal{O}(64{,}896)$ & $\mathcal{O}(64{,}896)$ & $\mathcal{O}(129{,}792)$ \\
   $4/9$   &  $\mathcal{O}(109{,}512)$ & $\mathcal{O}(328{,}536)$ &  $\mathcal{O}(657{,}072)$  \\
   $5/11$   &  $\mathcal{O}(490{,}776)$ & $\mathcal{O}(490{,}776)$ & $\mathcal{O}(981{,}552)$  \\
   $5/12$   &  $\mathcal{O}(48{,}672)$ & $\mathcal{O}(146{,}016)$ & $\mathcal{O}(292{,}032)$ \\
   $5/13; 6/13$  & $(\Z{13}\times\Z{13})\rtimes\PS4$ &  $(\Z{13}\times\Z{13})\rtimes\PS4$  & $(\Z{13}\times\Z{13})\rtimes(\PS4\times\Z2)$\\
   $5/14$  &  $\mathcal{O}(49{,}686)$ &  $\mathcal{O}(49{,}686)$ & $\mathcal{O}(99{,}372)$\\
   $7/18$  & $\mathcal{O}(27{,}378)$ &  $\mathcal{O}(82{,}134)$ & $\mathcal{O}(164{,}268)$ \\
   $9/26; 11/26$  & $(\Z{13}\times\Z{13})\rtimes\PS3$ &  $(\Z{13}\times\Z{13})\rtimes\PS3$ & $(\Z{13}\times\Z{13})\rtimes \D{12}$\\
   $\left\{14;16;17;19\right\}/39$   &  $\mathcal{O}(12{,}168)$ &  $\mathcal{O}(36{,}504)$ & $\mathcal{O}(73{,}008)$ \\
   $\left\{29;31;35;37\right\}/78$  & \hspace{0.2cm} $\Z3\times(\Z{13}\times\Z{13})\rtimes\PS3$ \hspace{0.2cm} & $\mathcal{O}(9{,}126)$ & $\mathcal{O}(18{,}252)$ \\
  \bottomrule[1pt]
\end{tabular}
\end{center}
\caption{\label{tab:groups}
Possibilities for unified flavor symmetry groups (or their order, if too large) for fractional values of $\delta$ in the phenomenologically allowed interval $\delta/\pi\in\left\{1.50, 1.66\right\}$. 
Other values for $\delta$ lead to groups even larger than the ones listed, with orders easily exceeding $\mathcal{O}(10^5)$.
Groups $\langle A,B, G^{\pm}_{1,2}\rangle$ and $\langle A,B,X^{\pm}_{1,2}\rangle$ are $\mathcal{O}(\infty)$ irrespective of the value of $\delta$.
Groups are stated modulo the unphysical global $\Z2$ factor generated by $G^{-}_{4}$.
These results have been obtained with GAP \cite{GAP4}.}
\end{table}

Remarkably, only angles that are quantized in units of $13$ or $13/2$, i.e.\ $\delta/\pi=\Z{}/26$, 
lead to groups of acceptable size. The fact that a fraction of $13$ gives rise to a small group may not be surprising, 
given that \T already contains such a divisor. What \textit{is} remarkable though, is that 
putting $\delta$ to $-5(6)\pi/13$ or $-9(11)\pi/26$ fits phenomenology well, while at the same time it 
give rise to the \textit{smallest possible} finite groups that one can achieve
with all those generators -- as small as the one obtained for $\delta=0$ or $\delta=\pm\pi/2$, cf.\ table~\ref{tab:groups}.
We will shortly come back to investigate the phenomenology of these particularly symmetric points.

\medskip

We now comment on these results.
The fact that it is impossible to unify the full non-trivial $\Z2\times\Z2$
symmetry of neutrinos with \T of the charged leptons into a single finite group
may seem like a drawback. 
However, this gives important directions for model building:
If one insists on keeping the full $\Z2\times\Z2$ for neutrinos, there 
is no other way than embedding the residual symmetries directly into 
a continuous group such as $\SU3$ or $\mathrm{G}_2$.
Quite appealingly, this would point to the fact that there 
might be no intermediate scales, or complicated finite groups, involved
between the flavor breaking scale and the low scale.

Alternatively, one may be fine with just keeping those parts 
of the residual neutrino symmetries that form finite groups with \T.
Depending on the choice of $\delta$, the unified symmetry then will be
one of the groups listed in table~\ref{tab:groups}.
However, even in the ``maximally finite-symmetric case'' $G_{\mathrm{max}}$,
there are four independent degrees of freedom in the neutrino mass matrix.
Somewhat ironically this means that by fixing $\delta$ to a geometric value, we are forced to liberate 
another degree of freedom elsewhere.
This implies that there will, in general, be model dependent corrections to the complex-TBM form 
which are not constrained by the residual symmetries.
We stress that those corrections do not have to be large, as is known from explicit models.
For example, in \cite{Altarelli:2005yp,Altarelli:2005yx} it happens that TBM-mixing is realized at leading order
despite the fact that there is only a \Z2 residual symmetry,
while in \cite{Ding:2011qt} leading order TBM appears even in the complete absence of residual symmetries.

Finally, the third and most radical alternative would be to revisit the charged lepton sector and search for 
symmetries beyond \T which are able to produce asymmetric textures, while being able to incorporate
the full residual symmetries of the neutrinos.

\section{Predictions of the maximally symmetric case}
\begin{figure}[t]
 \centering
   \includegraphics[scale=0.53]{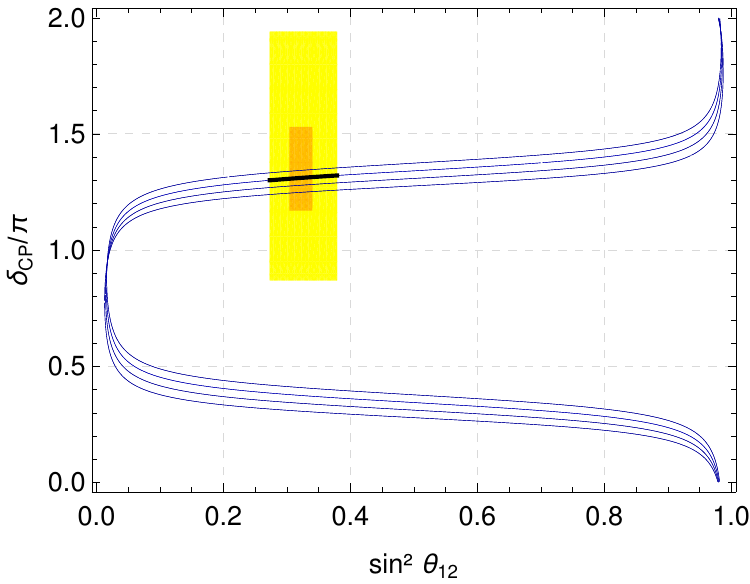}\hfill
   \includegraphics[scale=0.74]{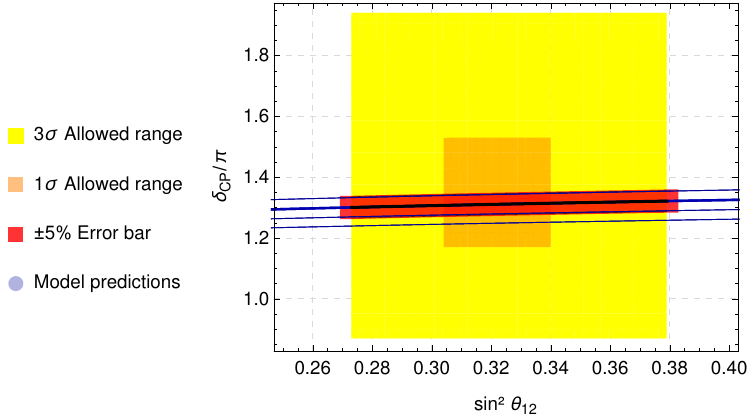} \\
  \includegraphics[scale=0.53]{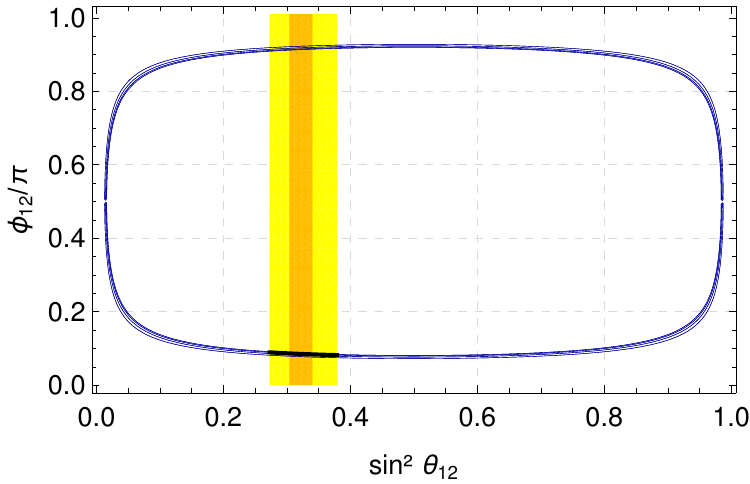}\hfill
  \includegraphics[scale=0.53]{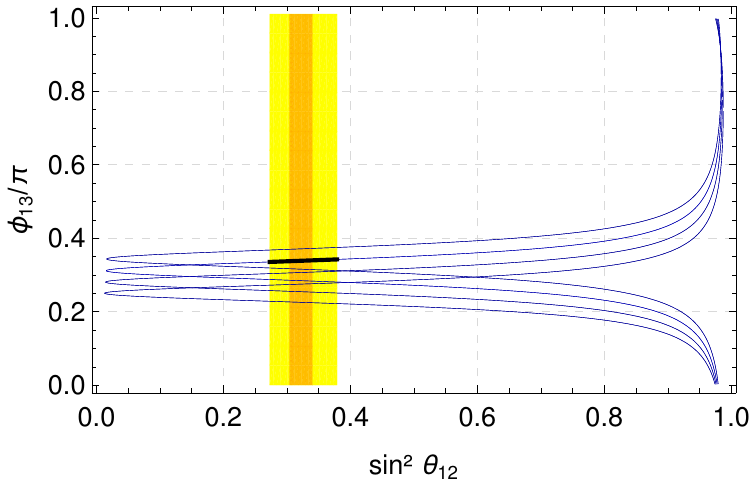}
  \caption{\label{fig:phasesvs12}
Prediction for CP violating phases in correlation with $\theta_{12}$ (blue lines) in the maximally finite-symmetric case, 
for fixed values of $\delta/\pi\times(-26)=\left\{9,10,11,12\right\}$ (the ordering is from bottom to top in the important regions of the plots). 
Highlighted in black is the range $\gamma\in\left\{-0.051,0.067\right\}$ for $\delta/\pi=-11/26$. 
On the top right is a zoom of the top left plot. There we show a $\pm5\%$ error bar for illustration on the $\delta/\pi=-11/26$ line, cf.\ the related discussion in the text. 
The related predictions for $\theta_{13}$ and $\theta_{23}$ are numbers, already stated in bold in table~\ref{tab:benchmark}.}
  \end{figure}

\begin{figure}[t]
 \centering
  \includegraphics[scale=1]{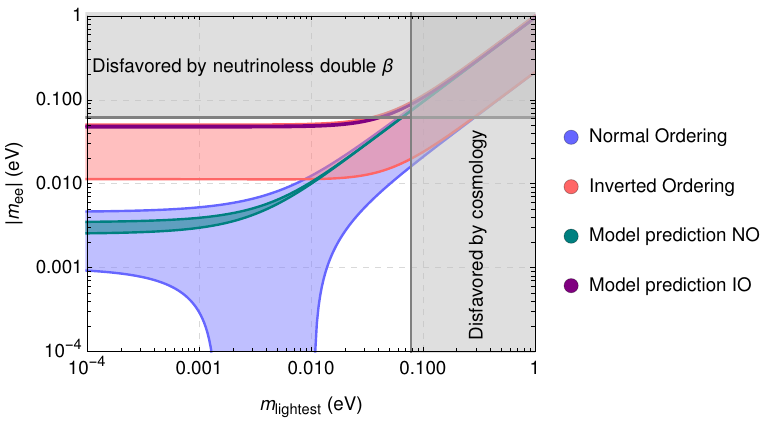}
  \caption{\label{fig:meeplot}
  Effective electron neutrino Majorana mass as a function of the lightest neutrino mass. 
  Varying the mixing parameters in their $3\sigma$ range leads to the blue(red) region for normal(inverted) ordering. 
  The predictions of asymmetric tri-bi-maximal mixing with $\delta/\pi=-11/26$, 
  as it may originate from a unified discrete flavor symmetry $(\Z{13}\times\Z{13})\rtimes \D{12}$, 
  are shown in green (NO) and purple (IO). We also show exclusion regions from Planck \cite{Aghanim:2018eyx} and 
  KamLAND-Zen \cite{KamLAND-Zen:2016pfg}.}
\end{figure}

We now focus on the maximally finite-symmetric case with unified flavor symmetry $G_{\mathrm{max}}$
and point out phenomenological consequences.
Taking $G_{\mathrm{max}}$ to be one of the two smallest possible unified flavor groups,
while requiring phenomenologically allowed mixing and CP violation, $\delta$ is restricted to take the values 
$\delta/\pi\times(-26)=\left\{9,10,11,12\right\}$.
For any value of $\delta$, the neutrino mass matrix that 
respects the residual flavor symmetry and GCP 
transformations $\langle G^{\pm}_{3}, X^{\pm}_{3},X^{\pm}_{4}\rangle$, takes the four-parameter form
\begin{equation}\label{eq:Mnu4}
 M_{\nu}~=~
\begin{pmatrix}
 a & b & -b\,\e{-\I \delta} \\
 b & c & \phantom{-}d\,\e{-\I \delta}\\
 -b\,\e{-\I \delta} & d\,\e{-\I \delta}  & \phantom{-}c\,\e{-2\I \delta}
\end{pmatrix}\;.
\end{equation}
Here, $a$, $b$, $c$, and $d$ are real parameters with dimensions of mass, 
and we go back to use our initial basis of eqs.~(\ref{eq:Ucl},\ref{eq:deltaTBM}).
The neutrino mixing matrix then is of the form\footnote{%
This can be shown using the analysis tools of \cite{Chen:2018zbq}. 
To convince oneself, one may also apply a rotation with $U_{\mathrm{TBM}\delta}$ to $M_{\nu}$ of eq.~\eqref{eq:Mnu4} and see that the missing step to a diagonalization is a $12$ rotation.}
\begin{equation}\label{eq:Udiagnu}
 U_\nu =U_{\mathrm{TBM}\delta}\,O_{12}(\gamma)\;,
\end{equation}
where $\gamma$ is one combination of the free parameters that corresponds to the angle of an 
orthogonal rotation $O_{12}(\gamma)$ in the $12$-block. 
The remaining three free parameters correspond to the neutrino masses.
Hence, also all column permutations of Eq.~\eqref{eq:Udiagnu} 
would diagonalize Eq.~\eqref{eq:Mnu4} with a viable mass spectrum.
However, $1\leftrightarrow2$ permutations, here, give rise to physically identical 
results (recall that $\theta_{12}$ can be adjusted by a free parameter).
By constrast, the other permutations are phenomenologically excluded 
since they cannot fit $\theta_{13}$ and $\theta_{23}$ at the same time. 
This implies that $\theta_{13}$ and $\theta_{23}$ remain fixed at their values of the original asymmetric texture, 
already stated in bold in table~\ref{tab:benchmark}.
In contrast, $\theta_{12}$ and all CP violating Dirac and Majorana phases become functions of the single free parameter $\gamma$. 
Hence, we can trade $\gamma$ for $\theta_{12}$ and the corresponding correlations are shown graphically in Fig.~\ref{fig:phasesvs12}.
To display Majorana phases we use the so-called symmetrical parametrization~\cite{Schechter:1980gr, Rodejohann:2011vc}
which is related to the standard parametrization~\cite{Tanabashi:2018oca} via $\alpha\equiv\alpha_{21}\equiv-2\phi_{12}$, $\beta\equiv\alpha_{31}\equiv-2\phi_{13}$, and 
$\delta_{CP}\equiv\phi_{13}-\phi_{12}-\phi_{23}$. The observational constraints on $\theta_{12}$ and $\delta_{CP}$ constrain $\gamma$ 
(for example, for $\delta/\pi=-11/26$) to the range $\gamma\in\left\{-0.051,0.067\right\}$; this fits well with the expectation of it being a small perturbation.
Our model assumes neutrinos with effective Majorana masses. 
Therefore, neutrinoless double beta decay is expected to be present. Since $\theta_{13}$ and $\theta_{23}$ are fixed while the CP phases are all strongly correlated this model has 
definite predictions for neutrinoless double beta decay. Fixing the lightest neutrino mass, the effective electron Majorana mass $m_{ee}$ only depends on one parameter. 
We show this in Fig.~\ref{fig:meeplot}. Clearly, this is a scenario in which also the case of normal 
mass ordering can be excluded by non-observation of neutrinoless double beta decay.

Altogether, the parameters obtained in this scheme for $\delta/\pi=-11/26$ and $\gamma=0$ are
\begin{align}\notag
s^2_{\theta_{13}}&\approx0.0220\,,&
s^2_{\theta_{12}}&\approx0.317\,,&
s^2_{\theta_{23}}&\approx0.496\,,& \\ \label{eq:predictions}
\delta_{CP}&\approx1.31\,\pi\,,& 
\phi_{12}&\approx0.09\,\pi\,,&
\phi_{13}&\approx0.34\,\pi\,,&  \\\notag
 m_{ee}^{\mathrm{NO}}&|_{m_{\mathrm{lightest}\rightarrow0}}\approx2.9\times10^{-3}\,\mathrm{eV}\,,& 
 m_{ee}^{\mathrm{IO}}&|_{m_{\mathrm{lightest}\rightarrow0}}\approx4.8\times10^{-2}\,\mathrm{eV}\,.&
\end{align}
These coincide with the values given in Tab.~\ref{tab:benchmark} for $\delta = -11 \pi/26$, since $\gamma = 0$. 
Varying the free parameter $\gamma$ will result in a modification of $\theta_{12}$ and the CPV phases 
while $\theta_{13}$ and $\theta_{23}$ preserve their values. The effect of such a variation is 
negligible at the current accuracy. For example, bringing $\theta_{12}$ to its current best fit value 
requires $\gamma\approx 4\times10^{-4}$.

Let us now comment on the precision of the predictions in our scenario.
First and foremost, we have neglected terms of $\mathcal{O}(\lambda^3)$ which roughly 
amounts to a $5\%$ uncertainty on the mixing angles. Also the values of $A$ and $\lambda$ itself 
are subject to uncertainties. 
Furthermore, one would expect corrections to the residual symmetry pattern originating from 
the RGE evolution below the flavor breaking scale. 
Assuming normal mass ordering and no new physics thresholds below the flavor breaking scale, 
the effect on mixing angles and phases is generically of the order $\mathcal{O}(10^{-5})$ \cite{Antusch:2003kp}
and, therefore, completely negligible.
For illustration, we show a $\pm5\%$ uncertainty band in figure \ref{fig:phasesvs12}.
The current level of precision makes it impossible even theoretically to differentiate
between the predictions of neighboring values of the fine-grained quantized angle $\delta$.

\section{Summary}

Recently, an \SU5 based model of grand unification with \T flavor symmetry was suggested by \cite{Rahat:2018sgs,Perez:2019aqq}. 
This very special flavor symmetry provides a connection between quark and charged lepton sectors, 
providing a symmetry origin for the so-called asymmetric texture of charged lepton mixing, eq.~\eqref{eq:Ucl}.
Together with the complex tri-bi-maximal mixing matrix for neutrinos, shown in \eqref{eq:deltaTBM}, 
this gives rise to a phenomenologically viable one-parameter ($\delta$) fit to lepton mixing and CP violation, see Fig.~\ref{fig:correlations}.

Here, we have studied how complex tri-bi-maximal mixing can be enforced by residual flavor and/or generalized CP symmetries of neutrinos, 
and how those symmetries can be unified with the \T flavor symmetry of the charged fermions.
Unification of the maximal residual symmetry of \eqref{eq:deltaTBM} with \T requires unified flavor symmetry 
groups of infinite order such as $SU(3)$ or $G_2$. 
However, if only parts of the maximal residual symmetry are imposed it is possible 
to unify them with \T to a finite unified flavor symmetry.
Reasonably small finite unified flavor groups are only obtained if $2\delta$ is quantized in multiples of $\pi/13$. 
In particular, $\delta/\pi=-11/26$ provides a good fit to the data while it can originate from the smallest possible 
unified flavor group $(\Z{13}\times\Z{13})\rtimes \D{12}$.

However, even though $\delta$ can be fixed by residual symmetries that form finite groups with \T, 
none of these groups is maximally restrictive on the neutrino mixing matrix. 
Thus, the residual symmetries allow for one additional unconstrained parameter, 
$\gamma$, which corresponds to a rotation in the 1-2 space. 
Nontheless, since $\gamma$ is observationally constrained to be a small perturbation, 
sharp predictions for CP violating phases and neutrinoless double-beta decay are possible, see eq.\ \eqref{eq:predictions} and Figs.~\ref{fig:phasesvs12} and~\ref{fig:meeplot}.

This model provides a compelling target for future experiments. It could be fully excluded
by a more precise measurement of $\theta_{23}$ or non-observation of neutrinoless double-beta decay -- even for normal neutrino mass ordering.


\begin{acknowledgments}
SCC would like to thank the Max-Planck-Institute for Nuclear Physics in Heidelberg for hospitality during his visit,
where this work was initiated. 
The work of SCC is supported by the Spanish grants SEV-2014-0398, FPA2017-85216-P (AEI/FEDER, UE) and PROMETEO/2018/165 (Generalitat  Valenciana), 
Red Consolider MultiDark FPA2017-90566-REDC and BES-2016-076643.
\end{acknowledgments}

\appendix
\section{Analytic expressions}
\label{app:AE}
Analytic expressions for the mixing angles are given by eq.\ \eqref{eq:th13} and 
\begin{align}
\sin{\theta_{12}}\,=&\,\frac{1}{\sqrt{3}}- \frac{\lambda}{3\sqrt{3}A}\left(A-2\cos\delta\right)-\frac{\lambda^2}{36\sqrt{3}A^2}\left(A^2-4A\cos\delta+4\cos2\delta\right)+ \mathcal{O}(\lambda^3)\;, \\ 
\sin{\theta_{23}}\,=&\,\frac{1}{\sqrt{2}}- \frac{\lambda^2}{36\sqrt{2}A}\left[A^2-4+4\left(A+9A^3\right)\cos\delta\right]+ \mathcal{O}(\lambda^3)\;.
\end{align}
The CP-odd Jarlskog-Greenberg invariant and the two Majorana invariants are given by 
\begin{align}
J\,=&\, \frac{\sin\delta}{9A}\left(\lambda -\frac{\lambda^2}{3}\right)+ \mathcal{O}(\lambda^3)\;,\\
I_1\,=&\, \frac{4\sin\delta}{9A} \left[\lambda - \frac{\lambda^2}{6A^2}\left(A-2\cos\delta\right)\right] + \mathcal{O}(\lambda^3)\;,\\
I_2\,=&\, \frac{4\sin\delta}{27A^2}\lambda^2\left(A+2\cos\delta\right)+ \mathcal{O}(\lambda^3)\;.
\end{align}
The Dirac and Majorana phases are given by 
\be
\begin{split}
 \sin\delta_{CP}\,=&\,\frac{2\sin\delta}{\left|A+2\e{\I\delta}\right|}\left[1-\frac{\lambda}{6A}\left(A+2\cos\delta\right)\right. \\
 & \left.+\frac{\lambda^2}{36A^2}\left(9 - A^2 + 36 A^4 + 8 (A + 9 A^3) \cos\delta + 11 \cos 2\delta\right)\right]+ \mathcal{O}(\lambda^3)\;,
\end{split}
\ee
\begin{align}
 \sin\alpha\,=&\,\sin\left(-2\phi_{12}\right)\,=\,\frac{2\sin\delta}{A}\left[\lambda+\frac{\lambda^2}{6A}\left(A-2\cos\delta\right)\right]+ \mathcal{O}(\lambda^3), \\ \notag
 \sin\beta\,=&\,\sin\left(-2\phi_{13}\right)\,=\,\frac{4\sin\delta}{\left|A+2\e{\I\delta}\right|^2}\left[\left(A+2\cos\delta\right)+\frac{\lambda}{6A}\left(A^2 + 4 A \cos\delta + 4 \cos 2\delta\right)\right. \\
 &\left.-\frac{\lambda^2}{36A^2}\left(A^3 + 2 A^2 \cos\delta - 4 A \cos2\delta - 8 \cos3\delta\right) \right] + \mathcal{O}(\lambda^3)\;.
\end{align}

\bibliographystyle{BibFiles/utphys.bst}
\bibliography{BibFiles/bibliography.bib}
\end{document}